\documentclass[structabstract]{aa}

\usepackage{graphicx}
\usepackage{amssymb}

\usepackage{natbib}

\usepackage{txfonts}
\usepackage{txfonts}

\usepackage{graphicx}
\usepackage{txfonts}
\usepackage{amsmath}
\usepackage{amssymb}

\begin{document}

\title{Testing an exact $f(R)$-gravity model at Galactic and local scales}
\author{{S. Capozziello\thanks{capozziello@na.infn.it}, E.
Piedipalumbo\thanks{ester@na.infn.it}, C. Rubano\thanks{rubano@na.infn.it}, \and P.
Scudellaro\thanks{scud@na.infn.it}} } \institute{{Dipartimento di Scienze Fisiche, Universit{\`a} Federico II di
Napoli and INFN - Sez. di Napoli, Complesso Universitario di Monte S. Angelo, Via Cinthia, Ed. 6, 80126 Napoli,
Italy } }
\date{Received \,/ Accepted \,}

\abstract {The weak field limit for a pointlike source of a
$f(R)\,\propto\,R^{3/2}$-gravity model is studied.} {We aim  to
show the viability of such a model as a valid alternative to GR
$+$ dark matter at Galactic and local scales.}{Without considering
dark matter, within the weak field approximation, we find general
exact solutions for gravity with standard matter, and apply them
to some astrophysical scales, recovering the consistency of the
same $f(R)$-gravity model with cosmological results.}{In
particular, we show that it is possible to obtain flat rotation
curves for galaxies, [and consistency with] Solar System tests, as
in the so-called "Chameleon Approach".  In fact, the peripheral
velocity $ v_\infty $ is shown to be expressed as $ v_\infty =
\lambda \sqrt{M}$, so that the Tully-Fisher relation is
recovered.}{The results point out the possibility of achieving
alternative theories of gravity in which exotic ingredients like
dark matter and dark energy are not necessary, while their
coarse-grained astrophysical and cosmological effects can be
related to a geometric origin.}


\titlerunning{Testing $f(R)$-gravity at Galactic and local scales}
\authorrunning{Capozziello S. et al}

\maketitle

\section{Introduction}

Alternative theories of gravity (Peebles and Ratra \cite{peebles};
Padmanabhan \cite{padmanabhan}; Copeland et al. \cite{copeland})
are increasing  as  possible suitable alternatives to dark energy
and dark matter. Although the $\Lambda$CDM model is affected by
many theoretical shortcomings (Carroll et al. \cite{carroll}),
and, in general, dark energy models are mainly based on the
implicit assumption that Einstein's General Relativity (GR) is the
correct theory of gravity. But the validity of GR both on large
astrophysical and cosmological scales still remains to be
accurately tested (see e.g. Will \cite{will}), and there is still
enough room to propose different theoretical schemes. A
\emph{minimal} alternative choice could be to take into account
generic functions $f(R)$ of the Ricci scalar curvature $R$. The
task for these extended theories is to fit the astrophysical data
without adding exotic dark ingredients (Kleinert and Schmidt
\cite{kleinert}; Capozziello \cite{capozziello1}; Capozziello et
al. \cite{carloni}; Odintsov and Nojiri \cite{odintsov};
Capozziello et al. \cite{capozziello1b}; Carroll et al.
\cite{duvvuri}; Allemandi et al. \cite{allemandi}; Nojiri and
Odintsov \cite{nojiri1}; Cognola et al. \cite{cognola1};
Capozziello and Francaviglia \cite{capozziello2}).

In such a context, these higher order theories have obtained
considerable attention in cosmology, since they seem to work well
both in the late and in the early universe (see Capozziello
\cite{capozziello1}; Capozziello et al. \cite{carloni}; Odintsov
and Nojiri \cite{odintsov}; Capozziello et al.
\cite{capozziello1b}; Carroll et al. \cite{duvvuri}; Allemandi et
al. \cite{allemandi}; Nojiri and Odintsov \cite{nojiri1}; Cognola
et al. \cite{cognola1}; Capozziello and Francaviglia
\cite{capozziello2}). It is also possible to show that $f(R)$
theories can play a major role at astrophysical scales, due to the
modifications of the gravitational potential
 in the low energy limit. Such a corrected potential reduces to the Newtonian one at the Solar System
scale and could also offer the possibility of fitting galaxy
rotation curves and galaxy cluster potentials without the need for
large amounts of dark matter (Capozziello et al.
\cite{capozziello3}; Milgrom \cite{milgrom}, Bekenstein
\cite{bekenstein}; Capozziello et al. \cite{capozziello4};
Capozziello et al. \cite{capozziello5}; Sobouti \cite{sobouti};
Frigerio Martins and Salucci \cite{frigerio}; Mendoza and
Rosas-Guevara \cite{mendoza}; Capozziello et al.
\cite{capozziello8}). However, extending the gravitational
Lagrangian may give rise to many problems. These theories may have
instabilities (Faraoni \cite{faraoni}; Cognola and Zerbini
\cite{cognola2}; Cognola et al. \cite{cognola3}), ghost-like
behavior (Stelle \cite{stelle}), and they still need to be matched
with data from the low energy limit experiments that are well
understood by GR.

In summary, adopting $f(R)$-gravity leads to interesting results,
first of all at cosmological and galactic scales, even if, up to
now, it has not been possible to select only \emph{one} theory (or
class of theories) good at all scales. There has been much work on
this (Capozziello et al. \cite{capozziello6}; Hu and Sawicki
\cite{hu}; Starobinsky \cite{starobinsky}; Nojiri and Odintsov
\cite{nojiri3}), but all the approaches are indeed
phenomenological and are not based on a fundamental conservation
or invariance principle of the theory.

In this paper, we propose a specific expression of the function
$f(R)$ of the Ricci scalar curvature $R$, namely\footnote{where
the  combination of minus signs is only due to our conventions,
since we start from the metric $(- + + +)$ and we obtain $R < 0$
and $G_{eff} = - 1/f^{\prime}$.}  $f(R) =-|-R|^{3/2}$, which comes
from  the need for the existence of a Noether symmetry for $f(R)$
cosmological models (de Ritis et al. \cite{deritis}; Capozziello
et al \cite{capozziello7}). The cosmological solutions of the
Einstein field equations related to such a choice for $f(R)$ have
been analyzed in Capozziello et al. \cite{capozziello7}, and it
turned out that this model admits a dust-dominated decelerated
phase, before a late time accelerated phase, as needed by the
observational data. Here, we study the low energy limit of such a
solution, in the case of a point-like source. We  consider the
Schwarzschild-like spherically symmetric metric in such a way
that, in the weak field limit, the Newtonian potential is modified
by adding a logarithmic term. A  similar treatment has been
proposed in Sobouti \cite{sobouti}, where instead the starting
point consists of introducing a specific hypothesis on the metric
and thereby deducing the form of $f(R)$ (resulting in a
power-law), so fitting observational data on speeds of peripheral
stars in spiral galaxies, as first reported in Sanders and McGough
\cite{sanders} and then selected in Sobouti \cite{sobouti}.
Unfortunately, this procedure leads to a very peculiar choice of
$f(R)$. It contains parameters which must be adjusted to the mass
of the gravitational source. Therefore, $f(R)$ cannot be
universal.

Here, we find results that are very similar to those in Sobouti
\cite{sobouti} from the observational point of view, but do not
exhibit the above problems. In this preliminary work, we only
consider the weak field generated by a point-like source. Clearly,
shrinking a whole galaxy (or cluster) to a point is  a very crude
approximation. Our aim, therefore, is to show that the model can
nonetheless work, without trying to obtain a strict correspondence
with observations.

The treatment of point-like source in $ f(R)$ theories is
non-trivial and, in a sense, can be considered an ill posed
problem. The reason is that we cannot disregard the properties of
the extended object that is  generating the field. Unlike what
happens in GR, the choice of the integration constants is not
necessarily independent of its peculiarities, like density,
equation of state etc. We should therefore solve the equations for
the inner metric and then match with the exterior. This procedure
is valid for a star but much less meaningful for a galaxy. In any
case, we do not expect to have a full prediction of their
functional dependence. In fact, also in standard GR the linear
dependence on mass of the constant is obtained only from the {\it
observational} link with Newtonian gravity. This link with
observations is precisely what is lacking in the case of stars,
and is only preliminarily studied here. Therefore, in the
following we retain the assumption of point-like source,
dedicating Sec. 6 to a deeper discussion on this point.

There is another limit of the analysis presented here, lying in
the weak field approximation assumed from the beginning. The
asymptotic Minkowskyian behavior cannot recovered and, most of
all, does not shed light on the singularities of the metric. It is
therefore not possible to say what are the modifications of a
black hole so generated.

In Sect. 2, we work out the basic equations of our model, and in
Sect. 3 we study peripheral speeds in spirals. In Sect. 4, we
discuss some tests in the Solar System and, in Sect. 5, we comment
on gravitational lensing and MOND. In Sect. 6 the connection
between the law for a pontlike source and an extended body is
briefly discussed. In Sect. 7, we give our conclusions.

\section{Weak field approximation}

Our  theory of gravity is determined by  the Action:
\begin{equation}
S=\frac{1}{2}\int{f(R)\sqrt{-g}d^{4}x}\,, \label{a1}
\end{equation}
where $R$ is the Ricci scalar and we define $f(R)=-|-R|^{3/2}$.

As said above, we consider a static, spherically symmetric metric,
which will differ from the standard Schwarzschild form, due to the
different starting equations. We thus set:
\begin{equation}
ds^{2}=B(r)dt^{2}-A(r)dr^{2}-r^{2}(d{\theta}^{2}-{\sin}^{2}\theta d\phi ^{2})\,, \label{a2}
\end{equation}
with $A(r)$ and $B(r)$ radial  functions to be determined through
the modified Einstein field equations.

It is important to observe that, as  the angular coordinates are
dimensionless, we also use a dimensionless distance, so that we
have:
\begin{equation}
r=\rho/r_{s}\,,
\end{equation}
where $\rho$ is the physical radius and $r_{s}$ is a suitable
scale. The choice of this scale is a delicate point: we could
decide to use a universal scale or one that is specific for the
particular situation. This second choice does not  affect the
universal character of our Action, but the situation will become
clearer in the following.

Proceeding in this way, all quantities are dimensionless,
including $R$ and the result for velocities.  In order to restore
the appropriate dimensions, we should therefore multiply by an
appropriate fundamental Action, i.e. some numerical multiple of
$\hbar$. As we are in  vacuum, this clearly does not affect the
equations and their solution. On the other hand, this constant is
of course not irrelevant and has an influence on the coupling with
the test particle (peripheral star or other object). In our case
this arbitrariness will be resolved by restoring the physical
units in an appropriate way when defining the observational
objects and fixing the constant $ C_3 $ (see below), with the
prescription that we should obtain ordinary Newtonian gravity at
small scales.

In a weak field, of course, the functions $A(r)$ and $B(r)$ are
practically identified by means of their corrections to what we
expect in the Newtonian case for astrophysical situations. First
trying to understand how we can modify the Newtonian potential, we
write the first function as $B(r)=1-2\epsilon
y(r)/r+O({\epsilon}^{2})$, where $\epsilon$ is a suitable small
parameter. Analogously, we also assume $A(r)=1+2\epsilon
x(r)/r+O({\epsilon}^{2})$. Being in a weak field, $r$ cannot
extend to infinity, but we must  have $2\epsilon y(r)/r\ll1$ and
$2\epsilon x(r)/r\ll1$.

We can obtain our results depending on $r$, as for instance:
\begin{equation}
R(r)=\frac{\epsilon(4x^{\prime}(r)+2ry^{\prime\prime}(r))}{r^{2}}+O({\epsilon}^{2})\,, \label{a3}
\end{equation}
where prime denotes a derivative with respect to $r$. It is also:
\begin{equation}
R_{tt}(r)=\frac{\epsilon y^{\prime\prime}(r)}{r}+O({\epsilon}^{2} )\,. \label{a5}
\end{equation}

To write equations, we have to vary the Action $S$ with respect to
the metric tensor, always remembering that we are studying an
approximate situation. We also need to note that
\begin{equation}
\label{a6}\frac{df}{dR} = \frac{3}{2}\sqrt{-\frac{4x^{\prime}(r) + 2ry^{\prime\prime}(r)}{r^{2}}}\sqrt{\epsilon}
+ O({\epsilon}^{3/2})\,.
\end{equation}
The master equation is:
\begin{equation}
\label{a7}\frac{df}{dR}\left(  \frac{3R_{tt}}{g_{tt}} - R\right)  + \frac{1}{2}f = 0\,,
\end{equation}
that is:
\begin{equation}
\label{a7'}0=\frac{\sqrt{-\frac{2x^{\prime}+ ry^{\prime\prime} }{r^{2}}}(-8x^{\prime}+
5ry^{\prime\prime})}{\sqrt{2}r^{2}} \,,
\end{equation}
while the trace equation is given by:
\begin{equation}
\label{a8}3{\Box R^{\prime}} + R\frac{df}{dR}- 2f = 0\,,
\end{equation}
which is equivalent to:
\begin{align}
  &  \left(  9\sqrt{\epsilon}\left(  4{x^{\prime\prime}}^{2} +
{y^{\prime\prime}}^{2} - 8x^{\prime}x^{\prime\prime\prime}-
8x^{\prime }y^{\prime\prime\prime}+ r^{2}
{y^{\prime\prime\prime}}^{2} + 4x^{\prime
\prime}(y^{\prime\prime}+ r y^{\prime\prime\prime})\right.
\right.
+\nonumber\\
&  - \left.  \left.  4rx^{\prime}y^{(4)} - 2ry^{\prime\prime}(2x^{\prime
\prime\prime}+ y^{\prime\prime\prime}+ r y^{(4)})\right)  \right)
\times\nonumber\\
&  \times\left(  4\sqrt{2}r^{4} \left(  -\frac{2x^{\prime}+ ry^{\prime\prime}%
}{r^{2}}\right)  ^{3/2}\right)  ^{-1} + O({\epsilon}^{3/2})=0\,.
\end{align}

Neglecting higher order terms and dividing the master equation by
${\epsilon}^{3/2}$ and the trace equation by ${\epsilon}^{1/2}$,
we finally find
\begin{align}
&
\sqrt{-\frac{2x^{\prime}+ry^{\prime\prime}}{r^{2}}}({-8x^{\prime}
+5ry^{\prime\prime}})=0\,,\label{a9}\\
&  4{x^{\prime\prime}}^{2}+{y^{\prime\prime}}^{2}-8x^{\prime}x^{\prime
\prime\prime}-8x^{\prime}y^{\prime\prime\prime}+r^{2}{y^{\prime\prime\prime}
}^{2}+4x^{\prime\prime}(y^{\prime\prime}+ry^{\prime\prime\prime})-\nonumber\\
&  4rx^{\prime}y^{(4)}-2ry^{\prime\prime}(2x^{\prime\prime\prime}
+y^{\prime\prime\prime}+ry^{(4)})=0 {}\,. \label{a10}%
\end{align}
Such equations appear  only  difficult to handle, since Eq.
(\ref{a9}) can be algebraically solved for $x^{\prime}(r)$ in
terms of $y^{\prime\prime}(r)$. Discarding the solution that
removes the denominator in Eq. (2.10), we are then left with
\begin{equation}
x^{\prime}(r)=(5/8)ry^{\prime\prime}(r)\,. \label{a11}
\end{equation}

Substitution of this expression and its derivatives in Eq.
(\ref{a10}) leads to the general exact solution:
\begin{align}
y(r) &  =\frac{1}{4}C_{2}r^{2}+C_{3}+C_{4}r-\frac{1}{4}C_{2}r\cos
(2C_{1})+\nonumber\\
&\frac{1}{24}C_{2}r^{3}\cos(2C_{1})+\frac{1}{4}C_{2}r\cos(2C_{1})\log
r+{}\nonumber\\
&  {}\frac{1}{4}\,\mathbf{i}\,C_{2}r\sin(2C_{1})++\frac{1}{24}\,\mathbf{i}%
\,C_{2}r^{3}\sin(2C_{1})-\nonumber\\
&\frac{1}{4}\,\mathbf{i}\,C_{2}r\sin(2C_{1})\log r\,,
\end{align}
where $C_{1}$, $C_{2}$, $C_{3}$, and $C_{4}$ are arbitrary
(complex) constants. If we limit ourselves to considering $C_{2}$,
$C_{3}$, and $C_{4}$ as real constants, we can understand that
posing $\sin(2C_{1})=\mathbf{i} \,\alpha$ and thus
$\cos(2C_{1})=\sqrt{1+{\alpha}^{2}}$ (with $\alpha \in\mathbf{R}$)
makes $y(r)$ real:
\begin{align}
y(r) &
=\frac{1}{4}C_{2}r^{2}+\frac{1}{24}C_{2}(-\alpha+\sqrt{1+{\alpha}^{2}
})r^{3}+C_{3}+\nonumber\\
&\frac{1}{4}r(-(C_{2}\alpha+C_{2}\sqrt{1+{\alpha}^{2}}
-4C_{4})+\nonumber\\
&  C_{2}(\alpha+\sqrt{1+{\alpha}^{2}})\log r)\,.
\end{align}
Here, the constant $C_{4}$ is multiplied by $r$ and, since the potential is obtained from $y(r)/r$, it gives
rise to a constant term. So, we can set $C_{4}=0$ in the following. On the other hand, the presence of a term
giving the Newtonian potential explicitly depends on $C_{3}$, which then needs to be nonzero.

From now on, $\epsilon$ will be incorporated into the integration
constants without any loss of generality.

\section{Peripheral speeds}

We are now in a position  to study  how the speed of a test star
behaves at the periphery of a \textit{pointlike} galaxy in our
model  (i.e. at large distances from a point source of great
mass), once it is subjected to the dimensionless potential
$\Phi(r) \equiv\frac{y(r)}{r}$. We  write the speed $v$ of a
generic test star in this potential. Remembering that the quantity
under the square root is dimensionless, we adjust the dimensions
by means of the light speed $c$. This will fix the numerical
values of the constants:
\begin{align}
v(r)\equiv &
c\sqrt{-r\frac{d\Phi}{dr}}=\nonumber\\&\frac{c}{2}\sqrt{\frac{-C_{2}
r(\alpha+\sqrt{1+{\alpha}^{2}})+
r(3+r(-\alpha+\sqrt{1+{\alpha}^{2}}))+12C_{3} }{3r}}\,. \label{b2}
\end{align}
The comparison with the Newtonian case is obtained when
$\alpha=C_{2}=0.$ The value of $C_{3}$ has to be
$C_{3}=\frac{GM}{c^{2}r_{s}}$. On the other hand, to get
increasing velocities we need $C_{2}$ to be negative, and we
define $C_{2}\equiv- {v_{0}}^{2} \beta/c^{2}$. Defining
$v_{0}\equiv\sqrt{12GM/r_{s}}$, we can then rewrite $v=v(r)$:
\begin{equation}
v(r)=v_{0}\sqrt{\frac{1+\beta r(3(\alpha+\sqrt{1+{\alpha}^{2}})+r(3+r(-\alpha
+\sqrt{1+{\alpha}^{2}})))}{12r}}\,, \label{b6}
\end{equation}
with $\alpha$ and $\beta$ being dimensionless constants. Another
(dimensional) constant is hidden in  $ r = \rho / r_s $. As said
in Sec. 1, the value of these constants should be linked to the
inner peculiarities of the body. Here, we try a much simpler
procedure, and, by means of suitable guesses, try to see if there
is a convenient choice that reproduces the observations. The
formulation of the solution so as to have dimensionless quantities
was very helpful.

The first assumption is $\alpha\gg1$, which clearly is arbitrary
and can be justified only \textit{a posteriori}.

 When $\alpha\gg1$,  the expression under the square
root reduces to $(1/12)r^{-1}+(1/4)\beta r+(1/2)\alpha\beta$. In
fact, we find a correction of what is usually expected in the
Newtonian case, since the first term is the Keplerian one, the
third gives a constant speed, and the second must stay small. The
reason is that, as said above, we must keep $2y(r)/r \ll1$. This
means that, even if $\alpha\gg1$, $\beta$ must be small. However,
it cannot simply be set to zero, for this would  also remove the
other correction term.

The second assumption is therefore that $\beta $, being so small
with respect to $ \alpha $,  we can safely neglect the increasing
term. A rough estimate of the relevant correction term for a
galaxy is:
\begin{equation}
v_{const}\equiv
v_{0}\sqrt{\frac{1}{2}\alpha\beta}=231\sqrt{\alpha\beta}\,\,km\,s^{-1}
\label{b7}
\end{equation}
(where we have considered $M=10^{10}M_{\odot}$, $r_{s}=5Kpc$, and
$G\equiv G_{N}$),  indicating that $\sqrt{\alpha\beta}\simeq1$.

Being dimensionless quantities, $ \alpha $ and $ \beta $ can be
universal  numbers or arbitrary functions of $GM/(c^{2}r_{s})$,
i.e. the only  number we can form with the physical quantities at
our disposal. But this is true only in the point-like source
model. Another dimensionless quantity that can be considered for
an extended body is of course its radius, again divided by $ r_s
$, and many others.

The third third assumption is thus that we consider $\alpha$ and $
\beta$ slowly varying with respect to these kinds of parameters.
In other words, we expect a difference to appear only when a very
large change of scale is considered, e.g. passing from the Galaxy
to the Solar System, which is discussed in Sec. 6.

The scale radius $r_{s}$, having only the function of fixing the
units, can be chosen arbitrarily and we  set it as a multiple of
the Schwarzschild radius, $r_{s}=KGM/c^{2}$,  $K$ being an
unspecified positive (large) number (if $K$ is kept fixed, then
$r_{s}$ is chosen according to the first issue (see above); if, on
the contrary, we want $r_{s}$ to be specific, we may  suitably
adjust the value of $K$. We shall see that this makes no
difference).

We thus get:
\begin{equation}
v^{2}(\rho)\sim\frac{6c^{2}\alpha\beta}{K}+\frac{GM}{\rho}+\frac{3c^{4}\beta }{K^{2}GM}\rho\,,\label{b8}
\end{equation}
that is, a constant term and a Keplerian one, plus one increasing
contribution which can be neglected for sufficiently small $\beta$
and $\rho$. This shows that the relevant parameter is the
combination $\alpha\beta/K$, while $3\beta/K^{2}$ must be
negligible.  We have thus reduced our problem to the determination
of a unique parameter $ C = \alpha \beta / K $. We want however to
stress that this simplification is due  to the above assumptions.
Moreover, it is clear that assumption three is more appropriately
applied to $ C $ instead of its separate elements.

Any attempt to modify the Newtonian law, on the Galactic scale,
has to cope with the justification of the Tully-Fisher empirical
relation. This obliges one to set the parameters of the modified
part as depending on the mass of the source, which is impossible
if one starts from a universal modification. On the contrary, it
is appropriate here, where the parameters are specific to the
problem. This is indeed a very good feature of our approach.
Moreover, we have complete freedom in the choice of the
dependance. A reasonable assumption is to set $ C=
\frac{\alpha\beta}{K}$ proportional to some power of the mass M,
made dimensionless by dividing by a suitable reference mass. Thus,
we define $\lambda M_{1}^{n}\equiv 3\alpha\beta/K$, where $n$ is a
pure number and $M_{1}$ is  referred to $10^{10}M_{\odot}$. (Such
a definition of $\lambda$ is the same as the one for the parameter
$\alpha$ in Sobouti \cite{sobouti}.

If we neglect the term linear in $\rho$, we find:
\begin{equation}
v^{2}(\rho)\sim\frac{1}{2}c^{2}M_{1}^{n}\lambda+\frac{GM}{\rho}\,,\label{b9}
\end{equation}
which coincides with the relevant part of the expression found in
Sobouti \cite{sobouti}. On the other hand, the constant term can
be written as:
\begin{equation}
v_{const}=211985\cdot M_{1}^{n/2}\sqrt{\lambda}\,\,km\,s^{-1}\,.\label{b10}
\end{equation}
A determination of $\lambda$ and $n$ can be found by fitting this
formula to the data on flat rotation curves from Sanders and
McGough \cite{sanders} as in Sobouti \cite{sobouti}, where a data
list of $31$ spirals is suitably selected and reported, including
radii, total masses, asymptotic orbital speeds, and velocity
curves for each  of those galaxies.  The fit is made with a
non-linear regression algorithm and, as expected, we obtain the
same result, as illustrated in Fig. 1.
\begin{figure}[ptb]
\includegraphics[width=0.5\textwidth]{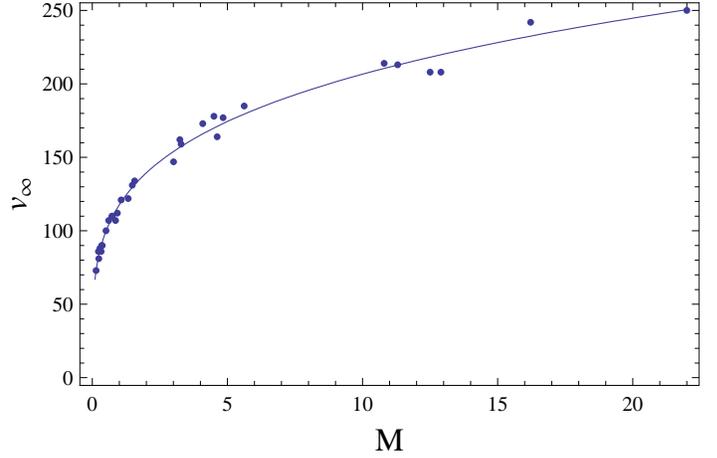}\newline\caption{Fit of Eq. (\ref{b10}) versus data in Sanders and McGough
\cite{sanders} as selected in Sobouti \cite{sobouti}. On the
$x$-axis, masses of luminous matter in units of
$10^{10}M_{\odot}$, peripheral speeds in $km\,s^{-1}$ in the
$y$-axis. Error bars are not indicated as they are not reported in
the reference papers. It is not  crucial, since we are discussing
only a preliminary estimate of the parameters.} \label{blob1}
\end{figure}
We obtain:
\begin{equation}
n=0.49 \pm 0.02\,\,\,\,\,\,\,\,\,\lambda=(3.1\pm
0.1)\cdot10^{-7}\,,\label{b10bis}
\end{equation}
compatible with $n=0.5$ at $1\sigma$. This yields:
\begin{equation}
v\,\,\propto\,\,M^{1/4}\,,\label{b11}%
\end{equation}
which is the phenomenological Tully-Fisher relation.  We have thus
proved that even in this crude model the flatness of the rotation
curves can be obtained (a more detailed verification would require
to treat the galaxy as an extended object); the immediate question
to answer is now the validity of the theory at the Solar System
level.

\section{Solar System tests}

\subsection{Kepler's law}

We now use the result of the above section for in the Solar
System, with the slight simplification $n = 1/2$. This is a major
extrapolation, as nothing guarantees that a semi-empirical law
like $v^{2}(\rho)\sim(1/2)c^{2}M_{1}^{n}\lambda+GM/{\rho}$ can be
used properly at such different scales, with the same values for
the parameters.  Possible arguments against this are:
\begin{itemize}
\item The power law guess may be not the right one. It could
indeed be more complicated and show  effects at small scales.

\item We do not know whether $\alpha\gg1$ at the Solar
System level or not.

\item The parameters used for the Solar System tests are those
used for galaxies, and they are already roughly estimated there.

\item The values used for $G$ and $M_{\odot}$ are the standard
ones, which is not certain, since they are  estimated within the
Newtonian framework.

\item The orbits of the planets around the Sun are only
approximately circular.

\item The Solar System is a many body system, where perturbations
should be accurately taken into account. (The observed deviations
from Kepler's laws are in fact explained by the existence of
perturbations induced by other planets.)

\item The passage from a pointlike source to an extended one is
 not trivial, as said above. This point seems the most important
 and is treated specifically below.
\end{itemize}

If we use Eq. (\ref{b9}) for the calculation of the Earth's
orbital speed, we find $v_{Earth} = 29.8\,km\,s^{-1}$, with an
error of $\sim1.4\,\%$, which is clearly large at this level.

In the Solar System we can also test  Kepler's third law
$T^{2}/{\rho}^{3}=C_{0}$ (with $T$ the period and $C_{0}$ a
constant), where:
\begin{equation}
T^{2}=\frac{4{\pi}^{2}{\rho}^{2}}{{v_{planet}}^{2}}\,,\label{c1}
\end{equation}
and if we use ${v_{planet}}^{2}=GM_{\odot}/\rho$, then
$C_{0}=4{\pi}^{2}/(GM_{\odot})$. When we instead use Eq.
(\ref{b9}) for ${v_{planet}}^{2}$, we find a non-constant result:
\begin{equation}
C(\rho)=\frac{4{\pi}^{2}}{GM_{\odot}+1.5\cdot10^{-7}c^{2}{M}_{1/2}
\rho}\,.\label{c2}
\end{equation}
The observed values of $C_{0}$ and those obtained from
Eq.(\ref{c2}) for  Solar System planets are plotted in Fig. 2
against the semimajor axes of the planet orbits. The deviation
from  observation is made evident by the absence of the downwards
trend, which is instead obtained from Eq. (\ref{c2}). However,
this is not a conclusive argument, as the trend could be altered
by the complicated effects of the many body system.
\begin{figure}[ptb]
\includegraphics[width=0.5\textwidth]{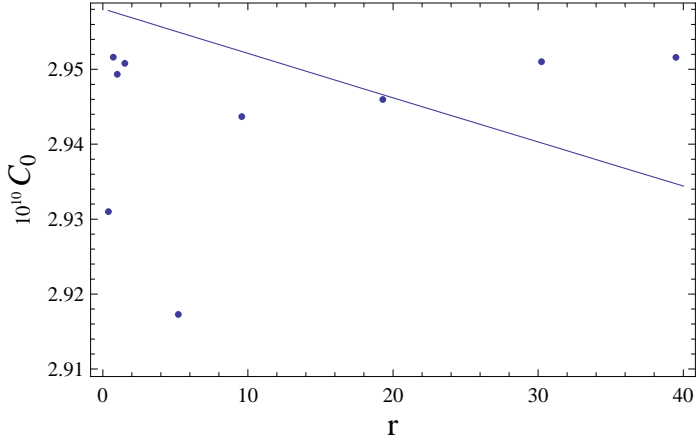}\newline\caption{Comparison between the values of the observed
$C_{0}$s for the planets of the Solar System and what is obtained
through Eq. (\ref{c2}). $x$-axis: the semi-major axes $\rho$ of
the planet orbits in the Solar System (in $AU$ units);  $y$-axis:
 the observed values of $C_{0}$ (\emph{points}) and the values of
$C(\rho)$ computed from Eq. (\ref{c2}) (\emph{line}).}
\label{blob2}
\end{figure}

\subsection{PPN parameter}

In the study of modified (with respect to GR) approaches to
gravitational physics, it is usual to ask which are the new values
for the PPN parameters, in particular of the most important and
studied $\gamma$, which must be very near to $1$ in the Solar
System (and exactly $1$ in GR).

Since we have developed our weak field approximation only up to
the first order, there is some doubt about the applicability of
the formalism. We have a modification of the $A(r)$ function in
the metric via the function $x(r)\neq1$, so that we may set:
\begin{equation}
\gamma=x(r)=1-\frac{5\lambda c^{2}\sqrt{M_{1}}}{4GM}\rho\,,
\label{d14}
\end{equation}
finding:
\begin{equation}
\gamma_{earth}=0.999\,.\quad\label{d15}
\end{equation}
Thus, $\gamma$ turns out to be $1$ up to $10^{-3}$, which is
indeed still far from the $10^{-5}$ approximation of the Cassini
experiment.  Moreover, we are not able to obtain other parameters,
due to the linear development in $ \epsilon $ of the metric.

The reason why we do not use the  conformal transformation to pass
to Einstein frame is that the equivalence can be lost in the case
of the weak field approximation due to the fact that a
$f(R)$-fourth-order theory has different gauge conditions in this
limit with respect to a second-order Einstein theory plus a scalar
field, which could lead to unphysical results. For a detailed
discussion of this point see Capozziello et al \cite{arturo}. In
order to avoid this, we adopted the Jordan frame for  all
calculations.

We obtain a value for $ \gamma $ which is not the same throughout
the Solar System. On the other hand, the numerical codes that give
an estimate of the PPN parameters are based on constant values. A
precise computation should instead take this feature into account.
We conclude that the transport of the correction as it is at the
Solar System level seems to be  unsatisfactory. In Sec. 5 this
problem will be addressed by the introduction of a "Chameleon"
mechanism.

\section{Other tests}

Still fixing, for simplicity, $n=1/2$,  we also set $K=1$ as it
introduces a constant term in the potential, which is clearly
irrelevant.

\subsection{Gravitational lensing}

Because it is intimately related to the underlying theory of
gravity in its Einstein formulation,  modifying the Lagrangian of
the gravitational field also affects the theory of gravitational
lensing. We therefore investigate how gravitational lensing works
in the framework of higher order theories of gravity. On the one
hand, one has to verify that the phenomenology of gravitational
lensing is preserved in order to not contradict those
observational results that do agree with the predictions of the
{\it standard} theory of lensing. On the other hand, it is worth
exploring whether deviations from classical results of the main
lensing quantities could be  detected and work as clear signatures
of a modified theory of gravity. As a first step towards such an
ambitious task, here we investigate how modifying gravity  affects
the gravitational lensing in the case of a point-like lens.

Indeed, the basic assumption in deriving the lens  equation is
that the gravitational field is weak and stationary. In this case,
the spacetime metric reads:
\begin{equation}
ds^2 = \left ( 1 + \frac{2 \Phi}{c^2} \right ) c^2 dt^2
\label{eq: weak}
\end{equation}
where $\Phi$ is the gravitational potential and, as usual, we have
neglected the gravitomagnetic term. Since light rays move along
the geodetics of the metric in Eq. (\ref{eq: weak}), the lens
equation may be simply derived by solving $ds^2 = 0$. Such a
derivation holds whatever  the theory of gravity is, provided that
one can still write Eq. (\ref{eq: weak}) in the approximation of
weak and stationary fields. As a fundamental consequence, the
lensing deflection angle will be given by the same formal
expression found in general relativity (Schneider et al.
\cite{SEF}; Petters et al. \cite{petters})
\begin{equation}
\vec{\alpha} = \frac{2}{c^2} \int{\vec{\nabla}_{\perp} \Phi dl}\,, \label{eq: vecalpha}
\end{equation}
with:
\begin{equation}
\vec{\nabla}_{\perp} \equiv \vec{\nabla} - \hat{e} (\hat{e} {\cdot} \vec{\nabla})\,, \label{eq: nablaperp}
\end{equation}
where $\hat{e}$ is the spatial vector\footnote{Here, quantities
 in boldface are vectors, while the versor will be denoted by
an over\,-\,hat.} tangent to the direction of the light ray and
$dl = \sqrt{\delta_{ij} dx^i dx^j}$ is the Euclidean line element.
The integral in Eq.(\ref{eq: vecalpha}) should be performed along
the light ray trajectory which is {\it a priori} unknown. However,
for weak gravitational fields and small deflection angles, one may
integrate along the unperturbed direction. In this case, we may
set the position along the light ray as:
\begin{equation}
\vec{r} = \vec{\xi} + l \hat{e}\,, \label{eq: defxi}
\end{equation}
with $\vec{\xi}$ orthogonal to the light ray.

Assuming the potential only depending on $r = | \vec{r} | = (\xi^2
+ l^2)^{1/2}$ (as for a point-like or a spherically symmetric
lens), the deflection angle reduces to
\begin{equation}
\vec{\alpha} = \frac{2 \vec{\xi}}{c^2} \int_{-\infty}^{\infty}{\left ( \frac{1}{r} \frac{d\Phi}{dr} \right )
dl}\,, \label{eq: alphacent}
\end{equation}
where we have assumed that the geometric optics approximation
holds, the light rays are paraxial and propagate from infinite
distance. Eq.(\ref{eq: alphacent}) allows us to evaluate the
deflection angle, provided that the source mass distribution and
the theory of gravity have been assigned, so that one may
determine the gravitational potential. Here, we consider only the
case of the point-like lens. Note that, although being the
simplest one, the point-like lens is the standard approximation
for stellar lenses in microlensing applications (e.g. see
Schneider et al. \cite{SEF}). Moreover, since in the weak field
limit $\vec{\alpha}$ is an additive quantity, the deflection angle
for an extended lens may be computed integrating the point-like
result weighted by the deflector surface mass distribution under
the approximation of a thin lens (i.e., the mass distribution
extends over a scale that is far smaller than the distances
between observer, lens and source) (Schneider et al. \cite{SEF}).
Given the symmetry of the problem, it is clear that we may deal
with the magnitude of the deflection angle and of the other
quantities of interest rather than with vectors.

In the approximation of small deflection angles, simple
geometrical considerations allow us to write the lens equation as
\begin{equation}
\theta - \theta_s = \frac{D_{ls}}{D_s} \alpha\,, \label{eq: lenseq}
\end{equation}
which gives the position $\theta$ in the lens plane of the images of the source situated at the position
$\theta_s$ on the source plane\footnote{Both $\theta$ and $\theta_s$ are measured in angular units and could be
redefined as $\theta = \xi/D_l$ and $\theta_s = \eta/D_s$ with $\xi$ and $\eta$ in linear units.}. Note that the
lens and the source planes are defined as the planes orthogonal to the optical axis, which is the line joining
the observer and the center of the lens. Here and in the following, $D_l$, $D_s$, $D_{ls}$ are the
observer\,-\,lens, observer\,-\,source, and lens\,-\,source angular diameter distances, respectively. In order
to evaluate the deflection angle, we need an explicit expression for the gravitational potential $\Phi$
generated by a pointlike mass. In our case the modified dimensional Newtonian potential, dropping unimportant
constant terms and with the above assumptions, is:
\begin{equation}
\Phi=\frac{GM}{\rho}-2c^{2}\lambda\sqrt{M_{1}}\log\left(  \frac{c^{2}\rho} {GM}\right)\,, \label{d1}
\end{equation}
where we should remember that $M_{1}$ is measured in units of $10^{10} M_{\odot}$.

The first term yields, of course, the usual deviation:
\begin{equation}
\delta_{0} \equiv \frac{4GM}{\xi c^{2}}\,, \label{d2}
\end{equation}
where $\xi$ is the angular impact radius. Computing the integral
$\int_{-\infty}^{+\infty}\nabla_{\bot}\Phi dl$ leads to the
additional deviation
\begin{equation}
\delta_{1} \equiv 2\pi\lambda\sqrt{M_{1}}\,, \label{d3}
\end{equation}
which is  independent of $\xi$. Now, in the case of the Sun we get
\begin{equation}
\delta_{0}\approx8.5\times10^{-6}\mbox{rad\quad;}\quad\delta_{1} \approx1.7\times10^{-11}\mbox{rad}\,,
\label{d4}
\end{equation}
so that the correction is irrelevant. On the other hand, in the
case of a galaxy, assuming $M_{1}=1$ and $\xi=10 Kpc$, we get
\begin{equation}
\delta_{0}\approx1.9\times10^{-6}\mbox{rad\quad;}\quad\delta_{1} \approx 1.7\times10^{-6}\mbox{rad}\,, \label{d5}
\end{equation}
so that the correction now dominates and induces an overestimation
of the mass, if the deviation is instead attributed in the usual
way\footnote{Here, we limit ourselves to observing that, of
course, the actual computation should be much more complicated,
not only because of the fact that a galaxy is an extended object,
but also because of the necessity to recompute the cosmological
angular diameter distances}.

The point-like lens equation  differs from the standard general
relativistic one for the second term (\ref{d3}). Should this term
be negligible, all the usual results of gravitational lensing are
recovered. It is therefore interesting to investigate in detail
how the corrective term affects the estimate of observable related
quantities since, should they  be detectable, they could represent
a signature of $R^{3/2}$ gravity. Since we are considering the
pointlike lens, a typical lensing system is represented by a
compact object (both visible or not) in the Galaxy halo acting as
a lens for the light rays coming from a stellar source in an
external galaxy, like the Magellanic Clouds (LMC and SMC) or
Andromeda. It is easy to show that, in such a configuration, the
standard Einstein angle $\theta_{E,GR}$ and the image angular
separation are of the order of a few ${\times} 10^{-5} \ arcsec$,
so that we are in the regime known as {\it microlensing} (see
Mollerach and Roulet \cite{MR02}).

In the \textit{standard} case the lens equation may be solved
analytically and one gets two images with positions given by
\begin{equation}
\vartheta_{{\pm,GR}} = \frac{1}{2} \left (\vartheta_s {\pm} \sqrt{\vartheta_s^2 + 4} \right ) \,. \label{eq:
poseinst}
\end{equation}
In the current case, the the lens equation may be conveniently
written as
\begin{equation}
 \vartheta^2 -\vartheta \vartheta_s - 1 =
 \frac{D_{ls}}{D_{s}}\frac{2\pi\lambda}{\theta_{E,GR}}\sqrt{M_1}\vartheta \,,
\label{eq: lenseqbis}
\end{equation}
with $\vartheta = \theta/\theta_{E,GR}$ and we have defined
$\theta_{E,GR}$ as the \emph{Einstein angle}, which in the general
relativity case is given by
\begin{equation}
\theta_{E,GR} = \sqrt{\frac{4 G m D_{ls}}{c^2 D_l D_s}} \,. \label{eq: einst}
\end{equation}

We still get two images on the opposite sides of the lens, with
one image lying inside and the other one outside the Einstein
ring. The geometric configuration is therefore the same as in the
standard case, but the positions are slightly changed. The images
positions are given by
\begin{equation}\label{imageposition}
\frac{1}{2} \left(A+\psi_\pm\sqrt{4+(A + \psi_s^2}\right)\,,
\end{equation}
with $A=2 \pi \lambda \sqrt{M_1}\frac{D_{ls}}{\theta_{E,GR} D_s}$.

To quantify this effect, we evaluate the percentage deviation
relative to the Einstein case.  $\theta_{E}=1/2 \left(A + \sqrt{4
+ A^2}\right)$, which differs from $\theta_{E,GR}$.

\subsection{Equivalence with MOND}

Following Sobouti \cite{sobouti}, we want now to show that our
model is effectively equivalent to the well known MOND theory (see
Milgrom \cite{milgrom}; Bekenstein \cite{bekenstein}), which
postulates a modification of Newtonian dynamics:
\begin{equation}
F=ma\mu\left(  \frac{a}{a_{0}}\right)  \,,\label{d6}
\end{equation}
where $\mu(x)$ is a suitable function, subject to the conditions:
\begin{equation}
\mu(x)=1~(x>>1)\quad;\quad\mu(x)\rightarrow0\ (x\rightarrow0)\,.\label{d7}
\end{equation}
$F$ is the usual gravitational force $F=GM/\rho$, while $a_{0}$ is
a universal parameter, that regulates the transition and is
estimated by means of a number of comparisons with observations,
with a value $a_{0}\approx1.2\times10^{-10}$ m/s$^{2}$. However,
the theory does not say anything about the form of the function
$\mu(x)$, which is highly degenerate.

In our case, we have that the asymptotically constant peripheral
velocity $v_{\infty}$ can be inserted into the formula for the
centripetal acceleration
\begin{equation}
a=\frac{v_{\infty}^{2}}{\rho}\,, \label{d8}
\end{equation}
so that using our expression in Eq. {\ref{b9}} for $v_{\infty}$,
we obtain
\begin{equation}
2a\rho=\lambda c^{2}\sqrt{\frac{M}{10^{10}M_{\odot}}}+\frac{2GM}{\rho}\,, \label{d9}
\end{equation}
from which we can extract $M$ and insert it into the expression of $\mu=\mu(a)$. (Of course, the MOND force
should now be interpreted as an effective one.)

There are two solutions, and we therefore obtain the functions:
\begin{equation}
\mu_{\pm}(a/a_{0})=\frac{8aG\times 10^{10}M_{\odot}+\lambda^{2}c^{4}\pm\lambda c^{2}\sqrt{16aG\times
10^{10}M_{\odot}+\lambda^{2}c^{4}}}{8aG\times 10^{10}M_{\odot}}\,. \label{d10}
\end{equation}

A first important remark is that $\mu$ is indeed independent of
the source mass as well as of distance, and it is therefore
universal, as requested. By comparison with the asymptotic limits,
on the other hand, we have to choose $\mu_{-}$ and get
\begin{equation}
a_{0}=\frac{\lambda^{2}c^{4}}{4G\times 10^{10}M_{\odot}}\,.\label{d11}
\end{equation}
From the MOND estimate of $a_{0}$ we then find
\begin{equation}
\lambda=2.8\cdot 10^{-7}\,,\label{d12}
\end{equation}
in very good agreement with the result extracted from the above
fit.

The great advantage of this formulation is that we obtain the universal
function:
\begin{equation}
\mu(x)=\frac{1+2x-\sqrt{1+4x}}{2x}\,, \label{d13}
\end{equation}
which can be tested.

\section{Point-like and extended bodies}

As discussed in previous sections, considering the corrections to
the potential from galaxies to Solar System scales, with the same
values of parameters, is rather unsatisfactory.  Here we want  to
discuss possible explanations and take into account also extended
bodies.

First, both the Sun and a galaxy are not point-like sources. In
the case of the Sun, however, assuming spherical symmetry and
neglecting as usual the rotational contribution, we may still use
the rotational invariant form of the metric in Eq. (\ref{a2}).
There are important differences if  the source is not point-like:
other elements may enter in the computation of the integration
constant, for instance the radius of the object. The correct
treatment would be to solve for the internal metric and join it
smoothly with the external one, so as to take care of the change
in density with radius. A reasonable approximation would be to
assume a uniform density, so that only radius could be involved.
Moreover, we may consider for the moment all stars as being equal
to the Sun, so that the correction to the potential can be
expressed again in terms of the mass
\begin{equation}\label{e1}
\Phi=\frac{Gm}{\rho}+\mu(m)\log(\rho/r_{s})\,.
\end{equation}
Here, $m$ is the mass of the star, $r_{s}$ is a scale radius, and
$\mu(m)$ is a function of the mass to be determined. Since we
consider all stars as being equal, $\mu$ is "universal" and
depends on the mass only.

The situation is very different for a spiral galaxy,  first,
because it is not spherical and second because it is not a
continuous distribution of mass. Thus, the correct procedure would
be to start from a metric with cylindrical symmetry, and to solve
the resulting fourth order equations. A very rough attempt would
consist of computing the force acting on a test particle near the
edge of the disk, by considering the galaxy as made up of, say,
$10^{10}$ stars like the Sun, (\textit{without either bulge or
intergalactic dust}, which is indeed very crude), and adding  the
forces.

The force acting on a unit mass test particle is therefore
\begin{equation}\label{e2}
F=\frac{Gm}{\rho^{2}}+\frac{\mu(m)}{\rho}\,.
\end{equation}

Let the test particle be situated on the x-axis, for instance (and
of course on the galactic plane), $R$ be the radius of the disk,
and $N$ the number of stars. The first term gives the usual
Newtonian force, so that we only have to sum  the correction. Due
to the non linearity of the dependence on the mass, this is not a
straightforward task. We give here just a rough estimate (a more
precise computation can be made and gives the same answer).

Because of the symmetry, the total force is clearly directed
towards the center, so that we need to compute only the x
component. The  correction is
\begin{equation}\label{e3}
\delta F=\sum\limits_{i=1}^{N}\frac{\mu(m)}{\rho_{i}}\cos(\theta_{i})\,,
\end{equation}
but, since $0<\rho_{i}<R$ and $0<\cos(\theta_{i})<1$, we may
substitute a mean value and write
\begin{equation}\label{e4}
\delta F=\sum\limits_{i=1}^{N}\chi\frac{\mu(m)}{R}=\chi N\frac{\mu(m)}{R}\,,
\end{equation}
where $\chi$ is a number of order 1. From the above discussion, on
the other hand, we have seen that a good empirical expression for
a galaxy is given by
\begin{equation}\label{e5}
 \delta F=\frac{\lambda_{\ast}\sqrt{Nm}}{R}\,,
\end{equation}
where $\lambda_{\ast}$  is the same as the $\lambda$ used before,
but in appropriate units. Comparing the two expressions we get
\begin{equation}\label{e6}
\mu(m)=\frac{\lambda_{\ast}\sqrt{Nm}}{\chi N}=\frac{\lambda_{\ast}\sqrt{m}%
}{\chi\sqrt{N}}\,.
\end{equation}

Remembering that $\chi\sim1$, we see that the force law for the
Sun is the same that we used for a galaxy, \textit{but with a
coefficient lowered by a factor} $\sim\sqrt{N}$. In other words,
the suppression of the correction acts in a twofold manner,
because of the very low mass of the Sun compared to a galaxy, and
because it is a single "almost point-like" object, instead of a
compound one. The consequence is that all the numerical values of
the local tests presented above should be lowered by a factor
$\sim10^{-5}$. In this case, the approach works very well.

The main problem of this argument lies not so much in the rough
estimate of $\chi$, but in the fact that  the number of stars is
not the same for all galaxies, and that the stars may be very
different in mass and size. A more satisfactory computation should
be done starting from Eq.(\ref{e2}), with a reasonable guess for
$\mu(m)$ (may be $\mu(m,r,...)$), and computing the rotation
curves with the right distribution of luminous matter and number
of stars. This is not a simple task and is postponed to future
work.

\section{Concluding remarks}

Starting from a reasonably simple $f(R)$ model for gravity, we
have shown that it is possible to obtain promising astrophysical
results, which do not require  dark matter. What is most important
is the fact that, with the  same model, it appears possible to do
the same also at a cosmological level, where exact solutions have
been discussed preliminarily in Capozziello et al.
\cite{capozziello7}. Of course, much work has still to be done on
both scales. The work here is only indicative of the concrete
possibility of testing a specific $f(R)$ model of gravity on
astrophysical grounds. For cosmology, at least the structure
formation and the cosmological microwave background radiation
spectrum must  still be investigated. On local astrophysical
scales, on the other hand, more realistic models for objects like
galaxies, for example, are necessary.

\end{document}